\documentclass[sigconf, nonacm]{acmart}
\usepackage{pvldb}

\renewcommand\vldbdoi{XX.XX/XXX.XX}
\renewcommand\vldbpages{XXX-XXX}
\renewcommand\vldbavailabilityurl{}

\usepackage[table]{xcolor}
\usepackage{multirow}
\usepackage{thmtools}
\usepackage{caption}
\usepackage{outlines}
\usepackage{subcaption}
\usepackage{tabularx}
\usepackage{array}
\usepackage[figuresright]{rotating}
\newcolumntype{Y}{>{\centering\arraybackslash}X}
\usepackage{booktabs}
\usepackage{hvfloat}
\usepackage{stmaryrd}
\usepackage[edges]{forest}
\usetikzlibrary{arrows.meta,shadows.blur, fit}
\usetikzlibrary{decorations.pathmorphing, calligraphy}
\usepackage{graphicx}      
\usepackage[normalem]{ulem}
\forestset{%
  colour me out/.style={
    draw=darkgray,
    rounded corners
  },
  rect/.append style={},
  dir tree switch/.style args={at #1}{%
    for tree={
      edge=-Latex,
      font=\ttfamily,
      fit=rectangle,
    },
    where level=#1{
      for tree={
        folder,
        grow'=0,
        s sep=1pt,  
        l sep=12pt,  
      },
      delay={child anchor=north},
    }{},
    before typesetting nodes={
      for tree={
        content/.wrap value={\strut ##1},
      },
      if={isodd(n_children("!r"))}{
        for nodewalk/.wrap pgfmath arg={{fake=r,n=##1}{calign with current edge}}{int((n_children("!r")+1)/2)},
      }{},
    },
  },
}

\usepackage{mathtools}
\usepackage[T1]{fontenc}
\usepackage{listings}
\usepackage[table]{xcolor}
\usepackage{tableaucolors}
\usepackage{algorithm}
\usepackage{algpseudocode}
\usepackage{amsmath}
\usepackage{amsfonts}
\usepackage{comment}
\usepackage{tabularx}
\usepackage{multirow}
\usepackage{soul}
\usepackage[normalem]{ulem}\usepackage[dvipsnames]{xcolor}
\usepackage[colorinlistoftodos,prependcaption,textsize=scriptsize,textwidth=1.2cm]{todonotes}
\usepackage{letltxmacro}

\LetLtxMacro{\origtodo}{\todo}
\renewcommand{\todo}[2][]{\origtodo[#1]{\raggedright\hspace{0pt}#2}}

\usepackage{amssymb}
\usepackage{enumitem}
\usepackage{multirow}

\usepackage{graphicx}
\theoremstyle{definition}
\usepackage{amsthm}

\makeatletter
\def\thm@space@setup{\thm@preskip=0pt
\thm@postskip=0pt}
\makeatother
\newtheoremstyle{newstyle}
{} 
{} 
{\mdseries} 
{} 
{\bfseries} 
{.} 
{ } 
{} 

\theoremstyle{newstyle}



\usepackage{hyperref}
\usepackage[most]{tcolorbox}
\newtcolorbox[auto counter, number within=section]{mybox}[2][]{colback=C7!10, colframe=C7, title={\textbf{#2}}, fonttitle=\small, title style={right}, top=0.5pt, bottom=0.5pt,boxsep=0.5pt, left=1pt, right=1pt, before=,after=, #1}
\newtcolorbox[auto counter, number within=section]{myboxNoTitle}[1][]{%
colback=C7!10,
colframe=C7,
top=0.5pt,
bottom=0.5pt,
boxsep=0.5pt,
left=1pt,
right=1pt,
before=,
after=,
#1
}
\lstdefinestyle{mystyle}{
    basicstyle=\small\ttfamily,
    stepnumber=1,
    numbersep=10pt,
    showstringspaces=false,
    breaklines=true,
    frame=single,
    rulecolor=\color{black},
    keywordstyle=\color{C0},
    commentstyle=\color{C7},
    stringstyle=\color{C3},
morekeywords={@prefix, a, sh:NodeShape, sh:targetClass, sh:property, sh:path, sh:qualifiedValueShape, sh:qualifiedMinCount, sh:class},
    literate={sh:}{{\textcolor{C2}{sh:}}}3
             {brick:}{{\textcolor{C1}{brick:}}}5
             {ex:}{{\textcolor{C5}{ex:}}}3,
    lineskip=-0.5ex, 
    aboveskip=1pt,
    belowskip=1pt,
}

\DeclareMathOperator*{\argmax}{arg\,max}

\algnewcommand\algorithmicinput{\textbf{Input:}}
\algnewcommand\Input{\item[\algorithmicinput]}
\algnewcommand\algorithmicoutput{\textbf{Output:}}
\algnewcommand\Output{\item[\algorithmicoutput]}

\makeatletter
\let\sout@orig\sout
\renewcommand{\sout}[1]{\ifmmode\text{\sout@orig{\ensuremath{#1}}}\else\sout@orig{#1}\fi}
\makeatother

\begin{document}
\title{Selective Retrieval Routing for Multi-Violation Knowledge Graph Repair}
\author{Tung-Wei Lin}
\affiliation{%
  \institution{UC Berkeley}
  \city{Berkeley}
  \country{USA}
}

\author{Gabe Fierro}
\affiliation{%
  \institution{Colorado School of Mines}
  \city{Golden}
  \country{USA}
}

\author{Han Li}
\author{Tianzhen Hong}
\affiliation{%
  \institution{Lawrence Berkeley National Laboratory}
  \city{Berkeley}
  \country{USA}
}

\author{Pierluigi Nuzzo}
\author{Alberto Sangiovanni-Vincentelli}
\affiliation{%
  \institution{UC Berkeley}
  \city{Berkeley}
  \country{USA}
}

\begin{abstract}

Knowledge graphs often contain missing or inconsistent facts that undermine the applications built on them. Schema validation detects such violations but does not identify which of several valid repairs recovers the intended graph. External corpora can supply the evidence needed to identify the intended repair, but retrieving from them is costly and not always successful. We propose a 
framework for selective retrieval in iterative graph repair. An LLM repair agent explores and edits the graph, while a router controls its access to a researcher agent that retrieves external evidence. We extend a violation-injection framework to generate cases with multiple compatible violations and known reference repairs. The agent's repair outcomes on these cases provide supervision for training the deployment-specific router. On two building-automation datasets, researcher access improves repairs for some violations and degrades them for others, with effects differing across deployments. The router uses fewer repair-time tokens than always enabling access in all six evaluation suites. It achieves higher repair $F_1$ than always disabling access in five of six suites.

\end{abstract}
\maketitle

\vldbtopmatter

\section{Introduction}\label{sec:intro}

Knowledge graphs (KGs) are an established abstraction for organizing heterogeneous data around entities and their relationships.
Public knowledge graphs such as DBpedia, Wikidata, and YAGO support information retrieval and other knowledge-intensive applications~\cite{auer2007dbpedia,vrandecicWikidataFreeCollaborative2014,suchanekYAGO45Large2024}.
In domain-specific operational settings, building ontologies such as Brick describe equipment, sensors, spaces, and their relationships so that analytics can be deployed across otherwise heterogeneous buildings~\cite{balaji2016brick,fierro2020mortar}.
Across these uses, the value of the graph depends on its quality: missing, stale, or inconsistent facts can cause queries and downstream applications to operate on an incorrect representation of the domain.

Such defects are difficult to avoid: KGs are typically assembled by integrating databases, tables, text, and other heterogeneous sources, often through imperfect automated extraction~\cite{melnyk2022knowledge,mihindukulasooriya2023text2kgbench}.
KGs can also drift out of date as the represented domain and its source artifacts evolve.
Graph constraint formalisms such as graph-generating dependencies~\cite{shimomura2020ggds}, SHACL~\cite{shacl_spec}, and PG-Schema~\cite{angles2023pg}
provide declarative specifications of the conditions that a valid graph must satisfy.
Validating a graph against these constraints identifies violations, but does not describe how to fix them.

Graph repair involves two problems: identifying repairs that restore schema validity, and selecting which of these valid repairs recovers the intended graph.
Symbolic reasoning and cost-based optimization techniques can compute valid repairs~\cite{arenas1999consistent,ahmetaj2022repairing,bohannon2005cost}, using minimality to choose among them.
This limits how much the graph changes, but does not establish which repair accurately represents the domain: the intended repair may require more edits than the smallest one.
Human-validated values and curated master data can resolve this ambiguity~\cite{fan2010towards,fan2019deducing}, but are limited by existing coverage and require human effort to maintain.
Recent methods also use large language models (LLMs) to inform candidate repairs, prompting models with graph context to propose corrections using knowledge acquired during pretraining~\cite{arnaout2022utilizing,terdalkar2025graph}.
Evidence can also be obtained directly from documents describing the modeled domain.
LLM-assisted graph completion and maintenance systems use retrieved passages to verify proposed triples or suggest replacements for outdated facts~\cite{zhou-etal-2024-cogmg,hatem2024uptodate}.
This provides a way to draw on existing knowledge corpora when choosing among schema-valid repairs.
However, the corpus is not guaranteed to contain the necessary facts, and querying it is not reliably successful.
Recent work on retrieval-augmented generation (RAG) shows that retrieval's benefit varies across tasks~\cite{asai2024self}, and that retrieved passages can mislead a model that would otherwise answer correctly~\cite{mallen2023not}.

We study an LLM graph-repair agent that explores the graph and synthesizes updates, with optional access to a \emph{researcher agent} that answers the repair agent's questions by retrieving from the external corpus; when access is enabled, the repair agent chooses whether to query the researcher and what to ask.

Our results show that the benefit of researcher access is \emph{heterogeneous}: it helps strongly for some kinds of violations, only marginally for others, and for some kinds it can actively \emph{hurt}, leading the agent toward worse repairs than it would have produced on its own.
This heterogeneity depends on the relationship among the constraints, the violation, the graph, and the deployment's retrieval pipeline.
When retrieval fails to improve a repair, the tokens spent querying the corpus add cost without improving quality.
Dynamically enabling researcher access for some violations and disabling it for others can therefore improve the tradeoff between repair quality and token cost.

This raises the question: \emph{when should an LLM graph-repair agent be given access to external knowledge?}
Hand-coding this decision requires knowing which violations benefit from retrieving  from the particular deployment's corpus; the answer can change when either the schema or the corpus changes.

We address this question by training a router that predicts, for each group of related violations, whether researcher access will improve the repair.
Since measuring repair quality requires knowing the intended repair, we construct test cases with known ground truth by introducing controlled sets of violations into a valid graph, varying the targeted predicates to obtain diverse cases.
We use these cases both to characterize researcher utility and to train the router.
We show how controllable multi-violation generation enables learning a deployment-specific access policy from a schema, a valid reference graph, and an accompanying corpus.
SHAP analysis~\cite{lundberg2017unified} of the learned router shows that the same predicate can increase the predicted benefit of access in one deployment and decrease it in another, illustrating how the learned predictions depend on the deployment.

We evaluate on two building-automation datasets with expressive SHACL constraints, instance graphs, and supporting document corpora (Sec.~\ref{sec:datasets}).
Our contributions are as follows:

\begin{itemize} [leftmargin=*]
    \item \textbf{Heterogeneous benefit of retrieval:} We show empirically that the effect of researcher access varies across violation types, improving some repairs and worsening others, even though the repair agent chooses whether to query the researcher.

    \item \textbf{Multi-violation test generation:} We extend the framework of~\cite{lin2025systematic} to generate test cases containing multiple compatible violations with known ground-truth repairs, varying the targeted predicates to produce diverse cases for evaluating repair agents and access policies.

    \item \textbf{Learned access routing:} We train a router on repair outcomes from generated cases to decide, for each group of related violations, whether to give the repair agent access to the researcher. In five of six evaluation suites, the resulting policy achieves higher repair $F_1$ than never granting access while using fewer repair-time tokens than always granting access.
\end{itemize}

\section{Related Work}

\label{sec:related}

Existing constraint-based repair methods search the space of possible repairs using cost-based optimization~\cite{bohannon2005cost}, probabilistic inference~\cite{rekatsinas2017holoclean}, and symbolic reasoning~\cite{ahmetaj2022repairing}.
A common repair criterion is minimality~\cite{arenas1999consistent}, which limits changes to the data but does not establish whether the resulting repair accurately represents the domain: a minimal repair may introduce incorrect information or remove correct information~\cite{fan2010towards,ahmetaj2022repairing}.
Semi-automated methods instead rely on additional evidence, such as curated master data with user-validated attributes~\cite{fan2010towards}, or user-validated values and entity matches~\cite{fan2019deducing}.
Because these methods use trusted evidence to identify correct repairs, their coverage depends on how complete that evidence is.
Our researcher agent serves a similar role, but consults an unstructured, multi-modal document corpus.
The corpus is not guaranteed to contain the necessary facts, and querying it is neither free nor reliably successful.

Another line of work learns repair patterns from previous corrections.
Edit histories can be used to mine correction rules~\cite{pellissier2019learning} or train a neural network~\cite{pellissier2021neural}.
LLMs offer a similar alternative: they have been applied to data wrangling~\cite{narayan2022can} and graph repair by treating the model's parametric knowledge as the evidence source~\cite{arnaout2022utilizing,terdalkar2025graph}.
These approaches depend on the availability of edit histories or relevant knowledge in the model; deployment-specific facts may instead require consulting external artifacts.
Related work on knowledge graph completion uses graph structure and parametric knowledge to propose candidate facts~\cite{wei2026kicgptv2,zhou-etal-2024-cogmg}.
Document-based verification and maintenance use external evidence to check proposed facts or suggest replacements for outdated ones~\cite{zhou-etal-2024-cogmg,hatem2024uptodate}.

Using external evidence also introduces a decision about when to retrieve it.
Selective RAG methods use signals about the model's knowledge, such as entity popularity~\cite{mallen2023not}, or assess the need for retrieval during generation~\cite{asai2024self,jiang2023active}.
Adaptive-RAG learns to select among no retrieval, single-step retrieval, and iterative retrieval based on query complexity, using automatically collected labels derived from model outcomes and dataset characteristics~\cite{jeong2024adaptive}.
Our work similarly learns from observed outcomes. In our setting, controlled graph corruption supplies reference repairs, and the router learns from outcomes along researcher-enabled and researcher-disabled repair trajectories. The learned policy controls access for each violation cluster within an iterative graph-repair process.

Our decision concerns whether to make a retrieval tool available to the repair agent, which, when access is enabled, chooses whether to query it and what to ask.
Tool-selection methods retrieve relevant tools from large catalogs~\cite{qin2024toolllm}; here, the available tool is fixed, and the question is whether access to it improves a particular repair.

Learning and evaluating this policy requires repair cases with ground truth.
Controlled error injection provides a way to generate such cases at scale; for example, BART~\cite{arocena2015messing} introduces constraint violations into clean relational data, with controls over error frequency, detectability, and repairability.
We reuse the derivation of individual violation-inducing edits from the graph violation-generation framework of~\cite{lin2025systematic}. Our extension samples and composes compatible expressions into multi-violation cases with targeted predicate coverage. The resulting cases support both the study of retrieval utility and training the access policy.

\section{Background and Problem Definition}

\label{sec:prelim}

We build directly on the abstract syntax introduced in~\cite{ahmetaj2025common} and the violation-generation framework in~\cite{lin2025systematic}.

\subsection{Graph Repair Task}\label{sec:repair_task}

We adopt the \emph{common graph} model~\cite{ahmetaj2025common,lin2025systematic}, where the \textbf{instance graph} is a set of \textbf{triples} relating nodes by predicates or assigning values to node keys. This common graph model generalizes both RDF~\cite{RDF} and property graphs~\cite{angles2023pg}. A \textbf{schema} $\mathcal{S}$ is a set of \textbf{constraints} $(\mathit{sel},\varphi)$: the selector $\mathit{sel}$ picks the nodes in the instance graph to which the constraint applies, each of which must conform to \textbf{shape} $\varphi$. A \textbf{violation} occurs when a node selected by $\mathit{sel}$ does not conform to $\varphi$. An instance graph with at least one violation is \textbf{invalid}; otherwise, it is valid. We instantiate constraints in SHACL~\cite{shacl_spec} and target RDF graphs for concreteness, but the formulation is language-agnostic and applies equally to ShEx~\cite{shex_spec} and PG-Schema~\cite{ahmetaj2025common}.

The atomic \textbf{edit operations} on an instance graph are the addition and removal of triples. 
A \textbf{repair} is a sequence of such operations that transforms an invalid graph into a valid one. For RDF graphs, repairs can be
expressed as SPARQL updates~\cite{SPARQL}.

Given an invalid instance graph and schema, the graph repair task is to produce a sequence of edits that recovers the intended graph.
Multiple repairs may restore schema validity, and the intended repair may require more edits than the smallest one or be one of several equally small alternatives.
In cyber-physical knowledge graphs~\cite{balaji2016brick,hammarRealEstateCoreOntology,rasmussen2020bot,bergmann2020semantic}, the intended graph represents a particular real-world system, so choosing among repairs may require evidence not represented in the graph or schema.

For evaluation, we use the instance graphs supplied by the dataset authors as reference graphs and assume their factual correctness.
A \textbf{test case} is produced by \emph{injecting} violations into a reference graph: applying recorded edit operations that make selected nodes violate constraints in $\mathcal{S}$.
Reversing these edits recovers the reference graph and provides a ground-truth repair for the test case.
Other edits may also restore schema validity; we measure repair accuracy by agreement with the reference edits and report schema validity separately (Sec.~\ref{sec:hetero}).
Sec.~\ref{sec:multi_VE} shows how to generate test cases with targeted predicates.

\begin{figure*}[t]
    \centering
    \includegraphics[width=\linewidth]{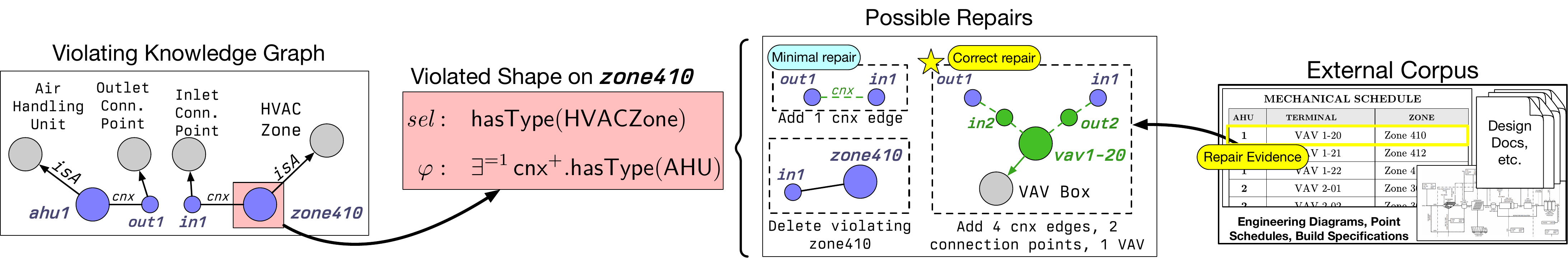}
    \caption{
        The illustrated shape requires each HVAC zone to reach exactly one air-handling unit through one or more \texttt{cnx} edges ($\texttt{cnx}^{+}$). Adding one edge between the existing outlet and inlet connection points satisfies this requirement. However, the equipment schedule identifies an intermediate variable air volume (VAV) box on the supply path. Recovering this documented path requires adding the VAV box and its connection points, yielding a larger (but correct) repair.}
    \Description{An incomplete HVAC graph, alternative repairs, and an equipment schedule identifying a variable air volume box between the air-handling unit and zone. The example contrasts adding an edge between existing connection points with recovering the documented intermediate equipment and connections.}
    \label{fig:repair_example}
\end{figure*}

\subsection{Study Setting and Datasets}

\label{sec:datasets}

We instantiate the graph repair task using two building-automation datasets modeled with ASHRAE Standard 223P~\cite{pritoni_marco_wetter_michael_paul_lazlo_prakash_anand_huang_weiping_bushby_steven_delgoshaei_parastoo_poplawski_michael_saha_avijit_fierro_gabe_steen_matt_bender_joel_ehrlich_paul_digital_2024}: the \textbf{Pritoni} dataset~\cite{pritoni2021metadata} and the \textbf{IBAL} dataset~\cite{pertzborn2024intelligent}.

We selected these datasets to satisfy two requirements. First, evaluating a retrieval-augmented system requires the LLM to not simply rely on its parametric memory. Benchmark contamination is well-documented in LLM evaluation~\cite{sainz2023nlp,balloccu2024leak}; if an agent has already memorized a widely referenced knowledge base like DBpedia~\cite{auer2007dbpedia} or Wikidata~\cite{vrandecicWikidataFreeCollaborative2014}, it bypasses the need for external retrieval entirely, confounding the evaluations. Because building-automation metadata is highly specialized, and because the Pritoni corpus is manually created by us, these datasets let us study retrieval of deployment-specific facts that are less likely to be available in the model's parametric knowledge.

Second, we require three artifacts describing the same system: a schema, a valid reference graph, and an external corpus that grounds the graph's facts.
To our knowledge, no existing benchmark provides all three.
Graph-repair benchmarks provide graphs and constraints but lack supporting documents~\cite{w3c_data_shapes_test_suite}; graph-construction benchmarks supply schemas and text but lack a reference instance graph to repair~\cite{mihindukulasooriya2023text2kgbench}; and question-answering corpora lack the graph and schema~\cite{mallen2023not,jeong2024adaptive}.
We therefore assemble these artifacts around the Pritoni and IBAL instance graphs, constructing a corpus for Pritoni and using actual deployment documents for IBAL.

Both the Pritoni and IBAL datasets are validated against the 223P schema, which models cyber-physical building systems by defining an \textit{ontology} to capture the physical and logical relationships between equipment and sensors. For example, the \texttt{observes} predicate links a sensor to its measured quantities (e.g., a room temperature).
Similarly, the \texttt{hasConnectionPoint} predicate links a piece of equipment to a port through which it exchanges a medium with other equipment (e.g., the supply-air outlet of an air-handling unit, joined by a duct to the inlet of a room).

Besides defining an ontology, 223P enforces 1{,}782 constraints on these relationships. For instance, a \texttt{minCount} constraint triggers a violation if a sensor node lacks \texttt{observes} relationships. Represented as an RDF graph, the 223P schema comprises 8{,}248 triples (expanding to 96{,}906 triples after resolving external imports). The majority of these triples define the underlying ontology and class hierarchies.
We now detail the instance graph and external corpus of each dataset.

\textbf{Pritoni Dataset.}
The Pritoni dataset models a single-story prototype building with five rooms spanning five distinct space types~\cite{pritoni2021metadata}. 
Its instance graph contains 1{,}707 triples over 384 entities.
Because the original release provided only the instance graph and the 223P schema, we manually constructed an accompanying external corpus to simulate real-world documentation: one network point list (a table of building-automation data points and their descriptions), two equipment schedules, four schematics (drawings and floor plans), and an 800-word HVAC sequence-of-operation document.
To reduce the risk of data leakage, this corpus was authored prior to violation injection and without reference to the agents' prompts or tools, limiting opportunities to tailor the documents to the sampled corruptions or agent configuration.

\textbf{IBAL Dataset.}
The IBAL dataset models a laboratory at the National Institute of Standards and Technology (NIST). Unlike Pritoni, IBAL is a real-world deployment and includes authentic engineering drawings and schedules. Its instance graph is roughly $3.5\times$ larger than Pritoni's, containing 6{,}016 triples over 1{,}357 entities. The accompanying external corpus comprises one network point list, 14 equipment schedules, and 16 engineering schematics.

We generate test cases from each dataset using the violation-injection framework of Sec.~\ref{sec:multi_VE}.
These repair cases are newly generated for this study, reducing the risk of evaluating on instances memorized during LLM pretraining.
The repair system must recover the reference graph, with the accompanying corpus available as an additional source of evidence. We next describe the repair system that operates on these test cases.

\section{Multi-Agent Repair System}\label{sec:agents}

In our system, an LLM \emph{repair agent} explores the instance graph and schema, synthesizes candidate updates, and uses validation feedback to revise them.
When the schema permits several valid repairs, the agent may need external evidence to recover the intended one.

Consider a graph that contains an air-handling unit and a zone in a heating, ventilation, and air-conditioning (HVAC) system, but lacks the supply path connecting them (Fig.~\ref{fig:repair_example}).
Adding a single \texttt{cnx} edge between the existing outlet and inlet connection points satisfies the illustrated shape with one triple addition, yielding a \emph{minimal repair}.
However, the accompanying equipment schedule identifies an intermediate variable air volume (VAV) box on the supply path.
Recovering the documented topology requires introducing the VAV box and its connection points, assigning their types, and adding the connecting edges.
Methods that minimize the number of edits would favor the one-edge repair over this larger reconstruction, illustrating why minimality alone does not identify the intended graph and why external evidence may be needed.

The repair agent can obtain evidence by consulting an optional \emph{researcher agent}, which searches the external corpus and answers the repair agent's questions.
When access is enabled, the repair agent chooses whether to query the researcher and what to ask, so having access does not mean that every repair involves retrieval.
Our router decides whether to enable this access for each \emph{violation cluster}, a group of related violations that are repaired together. We use the open-weight model \mbox{Gemma-4-31B}~\cite{team2026gemma} for all LLM calls.

\subsection{System Overview}

Our repair system addresses two practical challenges. First, real-world graphs may contain more violations than an LLM agent can reason about within its effective context window~\cite{liu2024lost}. Second, fixing violations in isolation may yield more incorrect repairs than resolving interacting violations jointly~\cite{chu2013holistic}. Thus, to satisfy LLM context limits while preserving interacting violation context, we group violations into clusters to resolve together. Finally, because a repair can ripple outward to resolve or create violations elsewhere, we revalidate the entire graph and recompute these clusters iteratively.

\begin{figure}
    \centering
    \includegraphics[width=\linewidth]{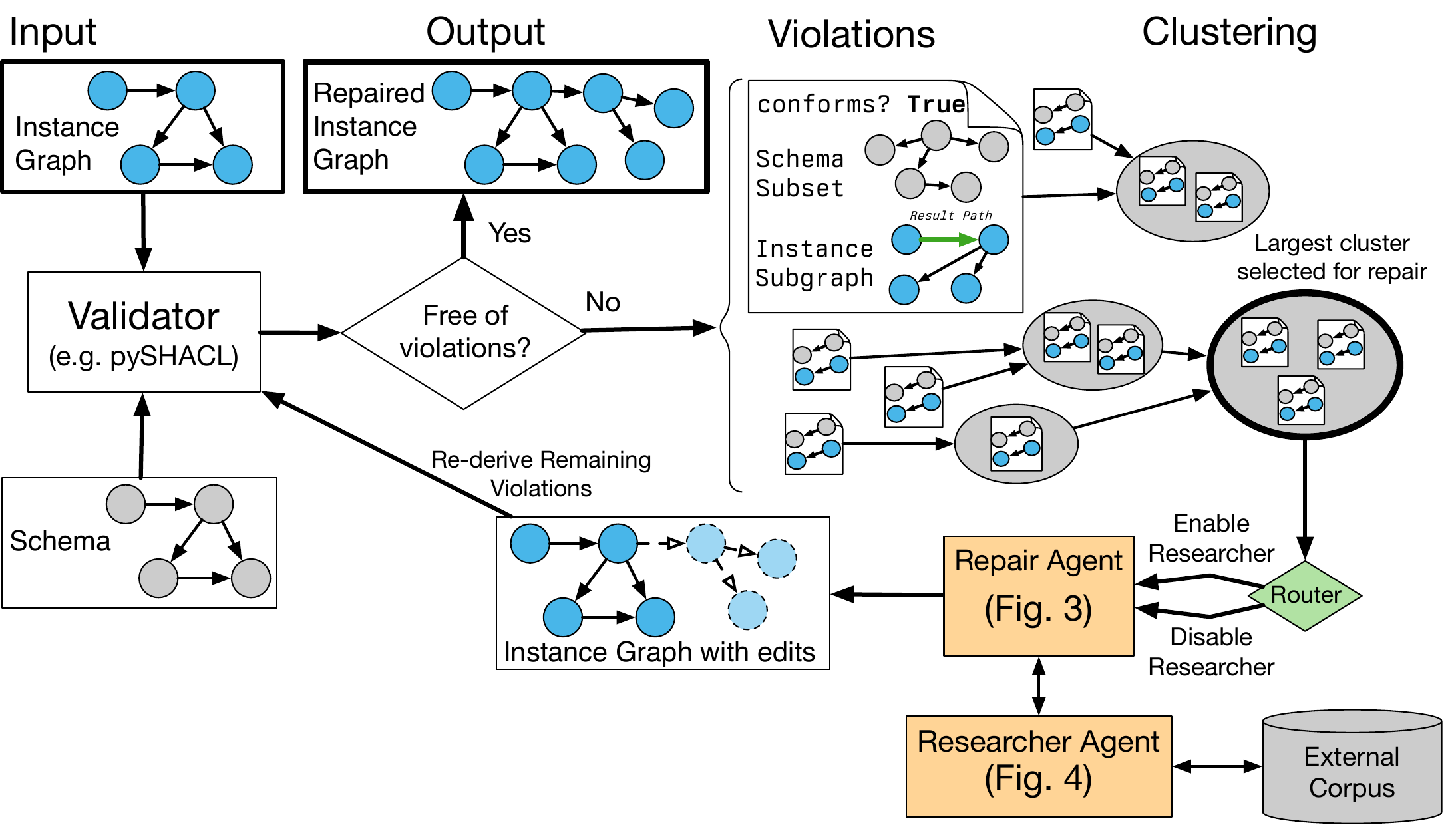}
    \caption{Overview of our iterative graph repair system, showing how violation clusters are provided to a repair agent, which optionally invokes a researcher agent to aid in generating a valid repair.}
    \label{fig:repair_loop}
    \vspace{-2em}
\end{figure}

As shown in Fig.~\ref{fig:repair_loop}, during each iteration, we first validate the instance graph against the schema to identify violations.
We then group related violations in the same cluster.
The system selects one cluster, synthesizes a repair for it, and discards the rest since updating the graph renders the remaining clusters stale.
After applying the repair to the full graph, we revalidate the updated graph to yield a fresh set of violations, initializing the next iteration.
When a repair attempt makes no progress, its violations are temporarily excluded from clustering and become eligible again after another repair makes progress.
The loop stops when the graph is valid or when it reaches five consecutive iterations without progress.

\begin{figure}[t!]
    \centering
    \includegraphics[width=\linewidth]{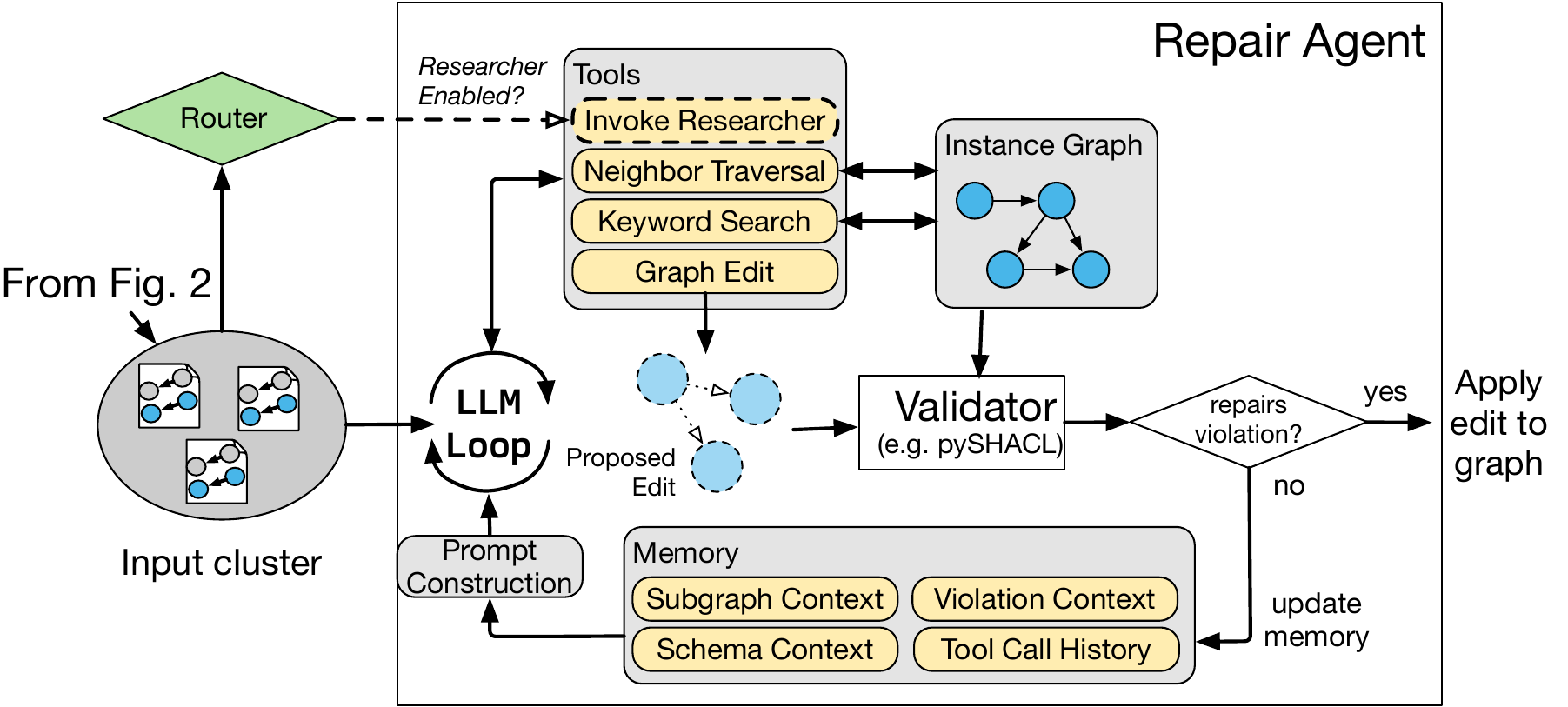}
    \caption[]{Iterative repair pipeline with agentic graph editing and SHACL validation. Graph exploration tools and the researcher agent also update memory through the tool-call history; these updates are omitted for clarity.}\label{fig:planner}
    \vspace{-2em}
\end{figure}

Our implementation has three components: a validator, a clustering algorithm, and a multi-agent repair component.

Recent systems like AutoGen~\cite{wu2024autogen} and MetaGPT~\cite{hong2024metagpt} demonstrate that delegating tasks across specialized agents reduces tool confusion and context pollution compared to monolithic agents.
Motivated by this, our system centers on a repair agent that explores and edits the graph and can optionally be equipped with a researcher agent that consults an external corpus. We evaluate the repair system under two prompt configurations. In the OFF configuration, the repair agent operates without access to the researcher and must rely solely on the graph and schema. In the ON configuration, the researcher is enabled, allowing the repair agent to ask the researcher natural-language questions whose answers are obtained from the external corpus. Sec.~\ref{sec:hetero} formalizes the corresponding experimental comparison.

Following Voyager~\cite{wang2023voyager}, we implement each agent as an iterative loop over structured memory rather than as a multi-turn conversation. At each step, we construct a fresh prompt containing the current state and a trace of the agent's prior tool calls, then ask the LLM to select the next action. We use two mechanisms to keep this prompt tractable during long reasoning chains. First, every four iterations, a separate LLM summarizes the tool-call history and replaces the execution logs in memory. Second, just before the thirty-iteration limit, we append a ``wrap-up'' instruction in the message to stop it from exploring and instruct it to commit to its best final action.

\subsection{Validator}\label{sec:validator}

We validate the instance graph using the SHACL validation engine pySHACL~\cite{pyshacl}, yielding a set of violations.
Each violation is augmented with two context payloads that will be used by the repair agent: a \emph{schema subset}, consisting of the failed constraint and its dependencies, and an \emph{instance subgraph}, containing the triples touched during validation of the failed constraint.
The clustering algorithm (Sec.~\ref{sec:clustering}) uses the instance subgraph to group violations, while the repair agent uses the schema subsets for local revalidation.

\subsection{Clustering and Batching}\label{sec:clustering}

We group violations into disjoint clusters to reuse shared context while limiting the context size with a budget $\tau$. Initially, each violation forms its own cluster. Merging two clusters takes the union of their instance subgraphs and the union of their schema subsets. Triples appearing in both clusters are included only once in the merged context, reducing the space needed to consider the violations together.

Formally, the clustering is based on two criteria: a \textit{size} budget and subgraph \textit{overlap}.
Because the context provided to the repair agent includes both the instance graph and schema information, the \textbf{size} of a cluster is defined as the combined size of its underlying unioned instance subgraph and schema subset.
We measure this size in nodes and set $\tau=20$ as a heuristic for these datasets; no individual violation exceeds this budget in our experiments.
We measure overlap using the Jaccard similarity of the instance-subgraph node sets, favoring clusters that concern the same graph region.
This increases the possibility of affecting multiple violations through a repair to the shared region.
Restricting the definition of overlap exclusively to the instance graph prevents clustering spatially distant violations merely because they violate the same schema constraints.

\noindent\textbf{Algorithm.}
Algorithm~\ref{alg:cluster} uses a greedy agglomerative heuristic. Every violation starts as its own cluster. We repeatedly merge the pair with the highest positive overlap among those whose combined size stays within $\tau$. We stop when no such merge remains, then pass the cluster containing the most violations to the repair agent. This procedure favors context sharing without claiming to find the largest feasible cluster.

\begin{algorithm}[t]
\caption{Constrained Agglomerative Clustering}
\label{alg:cluster}
\begin{algorithmic}[1]
\Input Violations $V$, size budget $\tau$
\Output A single cluster to pass to the repair agent
\State $\mathcal{C} \gets \{\, \{v\} : v \in V \,\}$ \Comment{each violation starts in its own cluster}
\Loop
    \State $\mathcal{P} \gets \{\, (c, c') \in \mathcal{C} \times \mathcal{C} : c \neq c',\ \mathrm{size}(c \cup c') \leq \tau \,\}$
    \If{$\mathcal{P} = \emptyset$ \textbf{ or } $\max_{(c,c') \in \mathcal{P}} \mathrm{overlap}(c, c') = 0$}
        \State \textbf{break}
    \EndIf
    \State $(c^\star, c'^\star) \gets \displaystyle\argmax_{(c,c') \in \mathcal{P}} \ \mathrm{overlap}(c, c')$
    \State $\mathcal{C} \gets \big(\mathcal{C} \setminus \{c^\star, c'^\star\}\big) \cup \{\, c^\star \cup c'^\star \,\}$ \Comment{merge best admissible pair}
\EndLoop
\State \Return $\displaystyle\argmax_{c \in \mathcal{C}} |c|$
\end{algorithmic}
\end{algorithm}

\subsection{Repair Agent}

As shown in Fig.~\ref{fig:planner}, the repair agent reads one cluster of violations and reasons about how to resolve them. In both ON and OFF configurations, the repair agent has access to the first three tools below (two graph exploration tools and one graph edit tool). In the ON configuration, it additionally has access to the researcher. Since both instance graph and SHACL schema are represented as RDF graphs, the agent uses the tools to explore both the instance graph and schema. Any attempt to edit the schema is rejected.

\begin{enumerate}[leftmargin=*]

\item Neighbor Traversal: Given a target node, return its immediate neighbors (i.e., the nodes pointing to or from it), optionally filtered by type, keyword, or predicate. This is used when the agent already knows the target node of interest.

\item Keyword Search: Given a keyword, return nodes with matching labels, optionally filtered by type. This is used when the agent has not yet located a node within the graph, for example, to match an entity mentioned by the researcher to its exact counterpart in the instance graph.

\item Graph Edit: The agent produces a repair as a SPARQL update, which is applied to the instance subgraph of the cluster and then automatically revalidated against the schema subset of the cluster. If the edited subgraph satisfies this schema subset, the loop exits and the repair is applied to the full graph; otherwise the agent continues, informed by the new validation outcome.

\item Researcher Agent (only for \textbf{ON} configuration): Query a researcher that consults the external corpus. Provisioning this tool makes retrieval available but does not force its use: the repair agent independently decides whether and when to invoke it. In the \textbf{OFF} configuration, the tool is omitted entirely. For example, given a node Channel\_78, the repair agent might ask the researcher: ``\textit{What property does Channel\_78 observe? Is it a temperature sensor?}''

\end{enumerate}

\subsection{Researcher Agent}

\begin{figure}[t!]
    \centering
    \includegraphics[width=\linewidth]{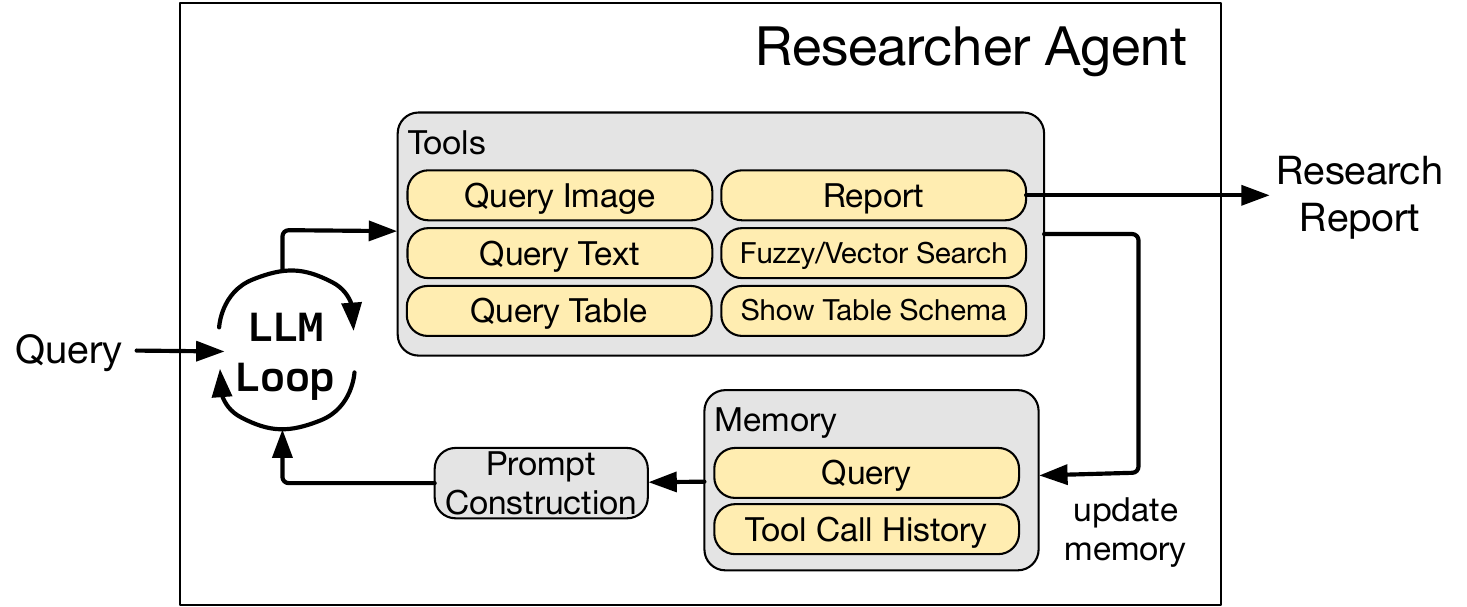}
    \caption[]{Iterative research pipeline with external corpus.}\label{fig:researcher}
    \vspace{-2em}
\end{figure}

As shown in Fig.~\ref{fig:researcher}, the researcher answers questions by searching an external corpus with multiple modalities: text, tables, and images (drawings, schematics, floor plans). It has the following tools:

\begin{enumerate}[leftmargin=*]

\item Term Search: Given an entity or keyword, the researcher queries a pre-computed index to locate relevant files. This index is constructed once prior to any retrieval by prompting an LLM to extract entities from images and by aggregating cell values from tables, enabling the agent to efficiently find where specific terms are mentioned.

\item Modality-specific readers: Once relevant files are identified, the researcher reads them in detail using a text reader, an image reader, or a SQL interface for tables, as appropriate.

\item Report: Concludes the research and returns the findings in natural language, e.g., it may answer: ``\textit{Channel\_78 (measurement\_id: ahu2\_heat\_rtd) observes temperature; it is a temperature sensor measuring the temperature downstream of the electric heater in AHU2.}''

\end{enumerate}

The researcher returns its report to the repair agent, which combines the report with its graph exploration to synthesize edits for the current cluster.

Beyond mitigating context bloat, this multi-agent architecture isolates the researcher as a modular tool, allowing us to control the repair agent's access to the external corpus. In the ON configuration, the repair agent’s prompt includes the researcher’s input-format description and usage instructions, making the researcher available as a tool. In the OFF configuration, these components are omitted, restricting the repair agent to graph exploration and editing. This toggle defines the ON and OFF conditions evaluated in Sec.~\ref{sec:hetero} and the binary action space of the routing policy in Sec.~\ref{sec:router}, which selects a prompt configuration for each violation cluster.

\section{Motivating Study: Retrieval is Heterogeneous}\label{sec:hetero}

The ON/OFF toggle determines whether the repair agent can consult the researcher.
We can therefore measure how access to the researcher changes repair outcomes.
Our results show that the utility of retrieval from an external corpus is not uniform across violations.
We organize the study around two factors: the repair's information requirement and the retrieval pipeline's efficacy.
These factors guide our predicate groupings and help interpret differences across deployments.
We therefore learn a deployment-specific routing policy from observed retrieval benefits, regardless of their underlying cause.

\paragraph{Information Requirement.}
Information requirement distinguishes repairs inferable from the invalid graph and schema (internal) from those requiring external evidence (external), following Paulheim's taxonomy on graph refinement methods~\cite{paulheim2016knowledge}.
The predicates involved provide an observable proxy for this requirement because they indicate what kind of information must be recovered.
For example, repairing a missing \texttt{observes} relationship requires identifying what a sensor measures---a physical fact that a network point list may supply.
Recovering a missing \texttt{hasConnectionPoint} relationship may instead require identifying an existing connection-point node through label matching and schema-based type filtering, using techniques similar to those in link discovery~\cite{ngonga2021limes}.
These differences motivate grouping predicates by their expected information requirements and testing whether retrieval utility varies across the resulting workloads.
The actual requirement also depends on the context available in the graph.

\paragraph{Pipeline Efficacy.}
Pipeline efficacy concerns whether the corpus and researcher together can supply useful evidence to the repair agent.
The kind of relationship expressed by a predicate also suggests how that evidence may be documented and recovered.
For example, a network point list may explicitly identify what a sensor measures, while recovering a connection shown in a schematic may require interpreting visual conventions and inferring the relationships they express.
Documents may describe physical equipment and connections without naming ontology constructs such as \texttt{ConnectionPoints}; using this evidence for repair also requires mapping it to the schema's representation.
Efficacy therefore depends on whether the deployment's documents contain sufficient evidence and whether the researcher can retrieve and interpret it.
We treat the corpus and researcher as a single pipeline whose output is the report available to the repair agent; Sec.~\ref{sec:limit} discusses the challenge of separating their contributions.
Differences in documentation and evidence recovery may therefore yield different retrieval benefits even for the same predicate, motivating comparisons across both predicate mixtures and deployments.

Retrieval utility depends on both what information a repair requires and whether the retrieval pipeline can supply useful evidence.
A repair may require facts absent from the graph, yet gain little from retrieval if the researcher cannot recover them.
Conversely, external evidence may help the agent even when the repair is inferable from the graph and schema.
These factors motivate studying retrieval utility empirically for each deployment.

\subsection{Instantiation in 223P}\label{sec:223p}

To study heterogeneity in retrieval utility in 223P, we group predicates into three broad classes based on where the information needed for repair is likely to be found: \emph{evidential}, \emph{terminological}, and \emph{representational}. We made this grouping by inspecting the 223P schema and considering what information is typically available in engineering documents. It provides a broad characterization of 223P rather than an exact taxonomy. Other ontologies may call for different groupings, and retrieval utilities may vary within a class across deployments.

\emph{Evidential predicates} describe relationships that are commonly documented in engineering artifacts, including point lists, equipment schedules, floor plans, P\&IDs, and sequences of operation. Examples include \texttt{observes}, \texttt{hasPhysicalLocation}, \texttt{actuates}, and \texttt{contains}. For example, an equipment schedule or floor plan may identify a sensor's location.

\begin{sloppypar}

\emph{Terminological predicates} are used to represent engineering concepts using the vocabulary defined by 223P and related ontologies.
Examples include \texttt{hasProperty}, \texttt{actuatedByProperty}, \texttt{hasDomainSpace}, and QUDT's \texttt{hasQuantityKind}.
Engineering documents often express these concepts through abbreviations, project-specific labels, or ordinary engineering language, which must be mapped to the terms available in the ontology.
\end{sloppypar}

\emph{Representational predicates} are used to construct relationships through structures required by the ontology. Engineering documents may describe a relationship without using the intermediate entities and links through which 223P represents it. The connection entities and points in Fig.~\ref{fig:repair_example}, linked by predicates such as \texttt{cnx}, \texttt{connectsThrough}, and \texttt{hasConnectionPoint}, illustrate this structure. These intermediate entities are similar to surrogate keys in a relational database: they are identifiers created by the data model to connect records. Repairing violations involving these predicates often requires constructing the intended structure from the schema and the surrounding graph.

We use these classes as a diagnostic lens for constructing test suites and examining how retrieval utility varies across predicate mixtures and deployments. They represent broad expectations about where repair information may reside.

\subsection{Measuring the Effect}

To measure the variation in retrieval utility, we use the violation injection framework to be introduced in Sec.~\ref{sec:multi_VE} to introduce violations into the valid Pritoni and IBAL instance graphs.
For each dataset, we generate three test suites, each predominantly targeting predicates from the evidential, terminological, or representational class.
Each combination of dataset and predicate mixture is a suite of $80$ test cases. A test case is also characterized by a parameter $K$  (the injection budget of the number of perturbations) and we distribute the suite over eight levels, $K \in \{2,5,10,15,20,25,30,35\}$, generating ten test cases at each level.
Sweeping $K$ lets us examine retrieval utility across repair tasks of varying size and complexity.

Given a test case and its ground-truth repair, we evaluate the repair system along three dimensions:

\begin{enumerate}[leftmargin=*]
\item $F_1$ Score: We compare the set of edit operations proposed by the repair system with the ground-truth edit set. The resulting $F_1$ score captures the precision and recall of the edits needed to reverse the injected perturbation.
\item {Schema Validity:} We record whether the graph produced by the repair system is valid, i.e., whether the edited graph has no remaining violations. We report this metric alongside $F_1$ because multiple repairs can satisfy the schema.
\item {Token Cost:} We measure repair-time token consumption, including all LLM calls for external research, summarization, and repair generation. 
\end{enumerate}

Every generated test case is evaluated under both ON and OFF configurations. In the ON condition, the researcher is available to the repair agent for every cluster; in the OFF condition, it is unavailable. The test case and all non-researcher components---including the repair loop and graph tools---are held fixed. Thus, the paired difference measures the effect of making the researcher available.
Comparing these differences across predicate mixtures and datasets characterizes variation in researcher utility.

The predicate mixtures provide a coarse way to vary the expected information requirement of retrieval utility. Evidential mixtures are likely to require external facts, whereas representational mixtures require mapping domain information onto 223P graph structures. However, the predicate class does not fully determine retrieval utility, which varies by individual predicate and deployment. For example, strong naming conventions in the graph may allow the repair agent to resolve an evidential violation through graph exploration; extracting implicit structure from point-label naming conventions is well established in building systems~\cite{bhattacharya2015automated, hong_building_2015, koh_scrabble_2018, waterworth_advancing_2021}. Moreover, even when external evidence is needed, retrieval helps only if the deployment's corpus contains the relevant information and the researcher can extract it. Retrieval utilities may therefore differ among predicates in the same class and across deployments.

To quantify this heterogeneity, we compute the paired difference for each test case.
This is defined as the subtraction between conditions.
For $F_1$ score, this is $\Delta F_1 = F_{1,\text{ON}} - F_{1,\text{OFF}}$.
We report the median paired difference along with a 95\% bootstrap confidence interval over the test cases (Fig.~\ref{fig:hetero}). Specifically, we resample 5,000 times with replacement from the paired differences pooled across $K$, with each resample the same size as the original set of paired cases. 
Token counts include every language model call: graph exploration, research, summarization, and repair synthesis.

To account for the eight injection-budget levels, we test whether the paired ON–OFF \(F_1\) differences tend to differ from zero, accounting for \(K\) when calculating the \(p\)-value for each dataset and predicate mixture (a \(K\)-stratified Wilcoxon signed-rank test~\cite{woolson2007wilcoxon}).


Before comparing $F_1$ scores, we note that across the 960 ON/OFF runs (80 test cases $\times$ 3 mixtures $\times$ 2 datasets $\times$ 2 configurations), only 6 fail to produce a schema-valid graph, a less than $1\%$ failure rate. In the following, we detail three observations regarding retrieval utility (Fig.~\ref{fig:hetero}):

\begin{figure*}
    \centering
    \includegraphics[width=\linewidth]{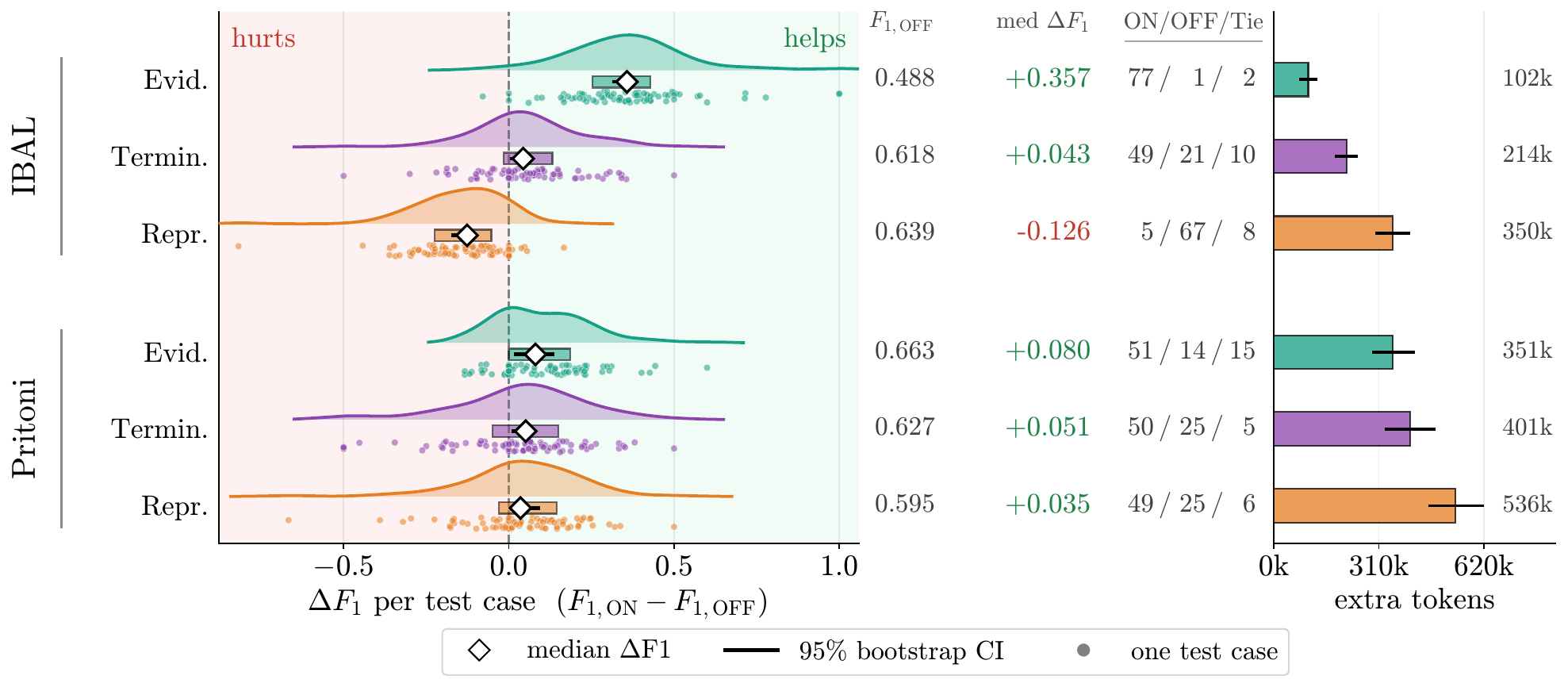}
    \caption{\normalfont\textbf{Impact of retrieval on repair accuracy ($\Delta F_1$) and token consumption across datasets and predicate mixtures.}~
Left: Paired differences ($\Delta F_1 = F_{1,\text{ON}} - F_{1,\text{OFF}}$) per test case, showing medians ($\diamond$) and $95\%$ bootstrap confidence intervals (bars). Reported two-sided $p$-values: on IBAL, Evidential ($p < 0.001$), Terminological ($p = 0.0016$), and Representational ($p < 0.001$); on Pritoni, Evidential ($p < 0.001$), Terminological ($p = 0.031$), and Representational ($p = 0.0036$). Right: Median extra tokens consumed by enabling retrieval (bar plots with $95\%$ CIs). For context, the median token consumption of the \textbf{OFF} condition: For IBAL, Evidential: $356\text{k}$, Terminological: $245\text{k}$, Representational: $296\text{k}$. For Pritoni, Evidential: $355\text{k}$, Terminological: $379\text{k}$, Representational: $426\text{k}$.
Normalized token efficiency ($\Delta F_1 / 100\text{k}$ extra tokens): IBAL yields $+0.352$ (Evidential), $+0.020$ (Terminological), and $-0.036$ (Representational), whereas Pritoni yields $+0.023$ (Evidential), $+0.013$ (Terminological), and $+0.007$ (Representational).}
\label{fig:hetero}
\end{figure*}

\textbf{Predicate mixtures reveal differences in retrieval utility.} On IBAL, enabling the researcher yields a median $\Delta F_1$ of $+0.357$ for the evidential suite, with 77 of 80 cases improving.
The terminological suite has a smaller median gain of $+0.043$, with both positive and negative case-level differences.
On Pritoni, the median gains are $+0.080$ (evidential), $+0.051$ (terminological), and $+0.035$ (representational).
The evidential suites have the largest median gains in both deployments, consistent with their emphasis on relationships commonly documented in engineering artifacts.
This pattern supports using predicates as observable proxies for differences in retrieval utility.
Pritoni's smaller evidential gain may reflect the context supplied by descriptive graph labels, which can help the repair agent infer physical facts through graph exploration.

\textbf{The effect of researcher access also depends on the deployment.} On IBAL's representational suite, the median $\Delta F_1$ is $-0.126$, with 67 of 80 cases getting worse.
On Pritoni's representational suite, the median gain is $+0.035$.
Thus, researcher access can help or hurt within the same predicate class.
Graph context, corpus coverage, and the researcher's ability to retrieve evidence offer possible explanations for this contrast.
The sign change motivates learning how predicates relate to retrieval utility within each deployment.

One observed repair trajectory illustrates a possible failure mechanism: a damper (\texttt{D14}) is missing a \texttt{hasConnectionPoint} predicate to its  \texttt{ConnectionPoint} node (\texttt{Damper\_14\_OCP}).
In the OFF condition, the repair agent uses a keyword search for ``Damper 14'' to identify the node \texttt{Damper\_14\_OCP}.
However, in the ON condition, the repair agent queries the researcher agent, which returns building descriptions (``the damper's inlet is a return-air duct and the medium is air'').
Following this report, the repair agent explores the graph using keywords ``return air duct'', ``air'', and ``RA''.
When these searches fail to locate \texttt{Damper\_14\_OCP}, the repair agent erroneously creates a new \texttt{ConnectionPoint} node to connect to the damper.

\textbf{Researcher availability has uneven returns.}
Normalizing median $\Delta F_1$ by the additional tokens consumed (Fig.~\ref{fig:hetero}) highlights an inefficiency. On IBAL, researcher availability yields an $F_1$ improvement of $+0.36$ per 100k additional tokens for evidential violations, compared to just $+0.02$ for terminological and $-0.04$ for representational violations. Similarly, on Pritoni, the return for representational violations is positive but small, at $+0.007$ per 100k additional tokens. Although the repair agent can choose whether to call the researcher under ON, it does not reliably determine when retrieval is worthwhile. A global ON policy therefore permits costly retrieval in regimes where its benefit is negligible or negative, whereas a global OFF policy forgoes retrieval where it is valuable. Because these regimes vary across predicates and deployments, researcher availability should be decided selectively.

\subsection{From Heterogeneity to Learned Routing}
These results motivate learning a deployment-specific routing policy.
The predicate classes organize our study of heterogeneity, but their aggregate effects differ across deployments: researcher access helps the representational suite on Pritoni and hurts it on IBAL.
We therefore use observed repair outcomes to learn access decisions within each deployment.

Effective training requires a representative set of test cases that cover the predicates the router will encounter at test time.
Our earlier experiments intentionally skewed the injected violations to reveal heterogeneity, but training a general router requires broad coverage.
We use inverse-frequency weighting across the full predicate set to encourage this coverage (Sec.~\ref{sec:pred_dist}); if a deployment's empirical violation distribution is known, that specific distribution can inform the sampling weights instead.
Next, Sec.~\ref{sec:multi_VE} introduces the sampler used to generate test cases, and Sec.~\ref{sec:router} presents the learned routing policy.

\section{Controllable Test Generation}\label{sec:multi_VE}
Our experiments require test cases with different predicate distributions. The suites in Sec.~\ref{sec:hetero} concentrate violations in the evidential, terminological, or representational predicate classes to expose retrieval heterogeneity. Router training instead requires broad coverage across predicates so that common predicates do not dominate the training data. Therefore, we extend the violation-generation framework of \cite{lin2025systematic} with a sampler that controls which perturbations are selected and composes multiple compatible perturbations into each test case.

Recall from Sec.~\ref{sec:prelim} that a test case is constructed by injecting violations. Each injected violation is represented by a set of graph edits that forces a node to break a constraint; reversing those perturbations provides the ground-truth repair. This section focuses on selecting and composing the perturbations.

\begin{figure}
    \centering

    \begin{tikzpicture}[
        >=Stealth,
        entity/.style={
            draw,
            ellipse,
            font=\ttfamily\small,
            minimum height=0.3cm,
            inner sep=1pt
        },
        relation/.style={
            ->,
            thick
        },
        edge label/.style={
            font=\ttfamily\small,
            fill=white,
            inner sep=1pt,
            outer sep=0pt
        }
    ]

    \node[entity] (node1) at (1.2,3.2)  {ahu\_1};
    \node[entity] (node2) at (3.6,2.6) {RAT\_Sensor};
    \node[entity] (node4) at (7.3,2.6) {``RAT\_01''};
    \node[entity] (node3) at (0.2,2.6) {AHU};

    \draw[relation]
        (node1.east) --
        node[pos=0.45, anchor=east, xshift=25pt, yshift=1pt,edge label] {hasPoint}
        (node2.west);

    \draw[relation]
        (node2.east) --
        node[pos=0.42, above=3pt, edge label] {description}
        (node4.west);

    \draw[relation]
        (node1.west) --
        node[pos=0.42, anchor=west,xshift=-13pt,yshift=3pt, edge label] {isA}
        (node3.north);

   \node[
    anchor=north,
    align=left,
    draw,
    rounded corners,
    fill=gray!10,
    text width=7.5cm,
    font=\small,
    inner sep=3pt
] at (4.,2.2) {
    \textbf{Schema $\mathcal{S}$:}\par
    \medskip
    \(\begin{aligned}
        \mathit{sel}:&\quad \texttt{hasType}(\texttt{AHU}),\\
        \varphi:&\quad
            {\exists^{\geq 1}}\,\texttt{hasPoint}.\texttt{NamedPoint},\\
        &\quad \text{where }\texttt{NamedPoint}
            \equiv
            {\exists^{\geq 1}}\,\texttt{description}.
            (\operatorname{len}\geq 3).
    \end{aligned}\)
};
    \end{tikzpicture}

    \caption{\normalfont\textbf{A valid instance graph example and its nested schema constraint.}~The point of \texttt{ahu\_1} has a description whose length is at least three, so \texttt{ahu\_1} satisfies the constraint.}
    \label{fig:ahu_schema}
\end{figure}
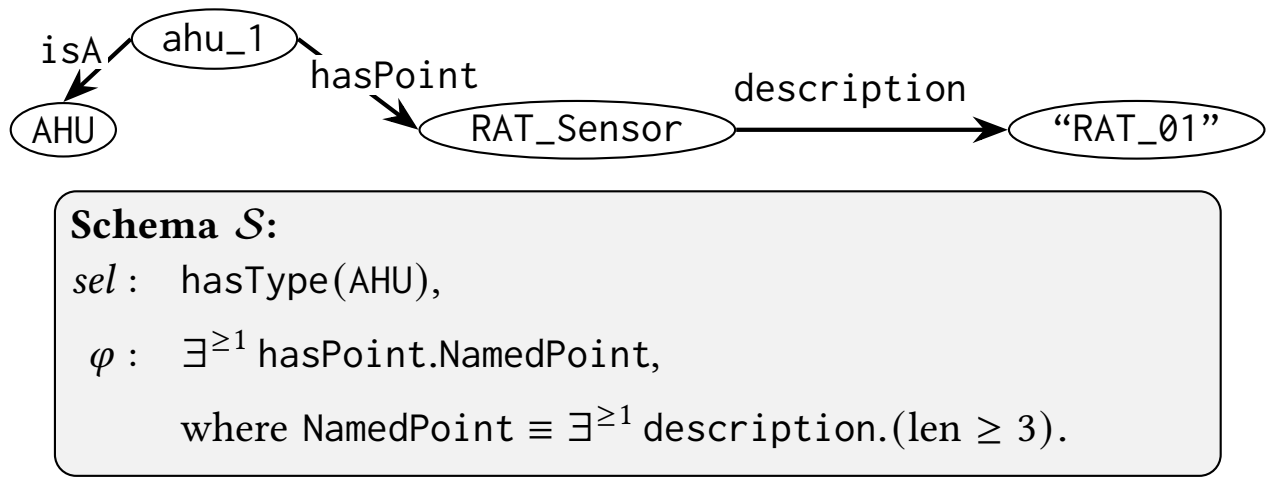

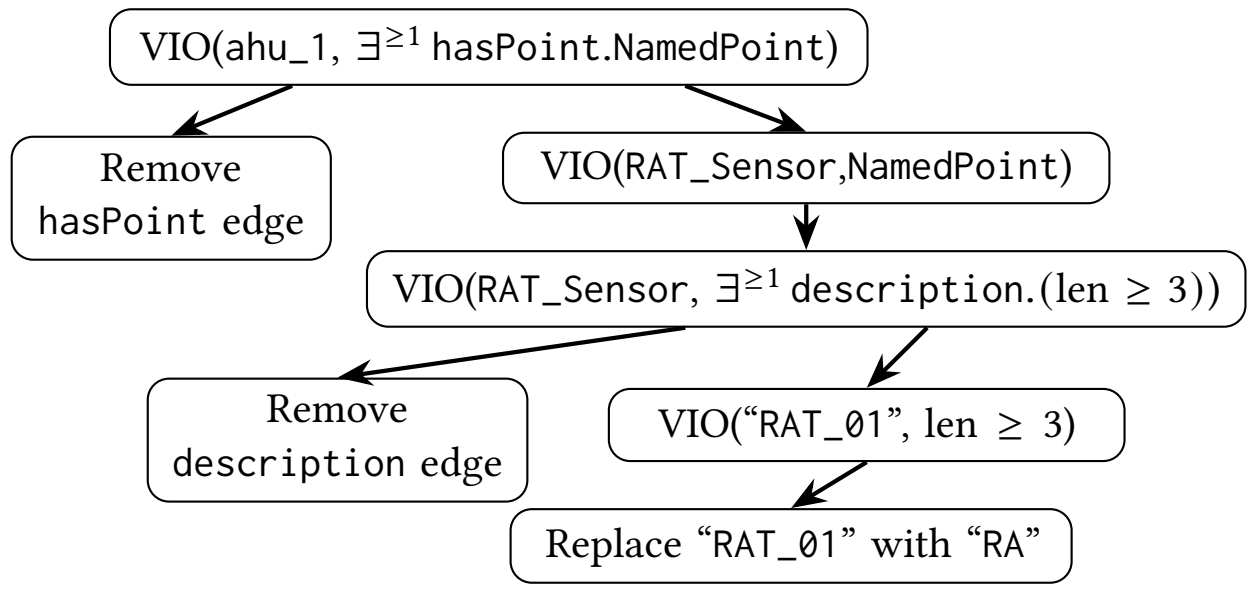
\begin{figure}
\centering

\begin{tikzpicture}[
    >=Stealth,
    repair/.style={
        draw,
        rounded corners,
        align=center,
        font=\small,
        inner sep=3pt
    },
    repair edge/.style={
        ->,
        thick
    }
]

\node[
    repair,
    text width=4.8cm
] (root) at (0,0) {
    VIO(\texttt{ahu\_1},
    \(
        {\exists^{\geq 1}}\,
        \texttt{hasPoint}.\texttt{NamedPoint}
    \))
};

\node[
    repair,
    text width=1.9cm
] (remove-point) at (-2.1,-1) {
    Remove\\
    \texttt{hasPoint} edge
};

\node[
    repair,
    text width=3.8cm
] (named) at (2.1,-0.8) {
    VIO(\texttt{RAT\_Sensor},\texttt{NamedPoint})
};

\node[
    repair,
    text width=5.6cm
] (sensor) at (2.1,-1.6) {
    VIO(\texttt{RAT\_Sensor},
    \(
        {\exists^{\geq 1}}\,
        \texttt{description}.
        (\operatorname{len}\geq 3)
    \))
};

\node[
    repair,
    text width=2.3cm
] (remove-description) at (-1,-2.6) {
    Remove\\
    \texttt{description} edge
};

\node[
    repair,
    text width=3.2cm
] (literal) at (2.5,-2.5) {
    VIO(``\texttt{RAT\_01}'',
    \(
        \operatorname{len}\geq 3
    \))
};

\node[
    repair,
    text width=3.5cm
] (replace) at (2.,-3.3) {
    Replace ``\texttt{RAT\_01}''
     with ``\texttt{RA}''
};

\draw[repair edge]
    ([xshift=-1.3cm]root.south) --
    (remove-point.north);

\draw[repair edge]
    ([xshift=1.3cm]root.south) --
    (named.north);

\draw[repair edge]
    ([xshift=-0.8cm]sensor.south) --
    (remove-description.north);

\draw[repair edge]
    ([xshift=0.8cm]sensor.south) --
    (literal.north);

\draw[repair edge]
    (literal.south) --
    (replace.north);
\draw[repair edge]
    (named.south) --
    (sensor.north);
\end{tikzpicture}

\caption{The expansion graph of violating the constraint $\varphi$ in Fig.~\ref{fig:ahu_schema}. Arrows represent alternative violation strategies.}
\label{fig:expansion}
\end{figure}

\subsection{Constructing Test Cases}

We first illustrate how candidate perturbations are derived from a valid graph and its schema. Fig.~\ref{fig:ahu_schema} shows an AHU node \texttt{ahu\_1} with a point \texttt{RAT\_Sensor}. The schema requires every AHU to have at least one \texttt{hasPoint} value conforming to \texttt{NamedPoint}; in turn, a \texttt{NamedPoint} must have a description whose length is at least three. The graph satisfies this constraint because \texttt{RAT\_Sensor} has the description
\texttt{"RAT\_01"}.

Following \cite{lin2025systematic}, an \textbf{expansion graph} represents alternative ways of making a selected node violate a constraint. It is a search structure constructed from the schema and instance graph, not a subgraph of the instance graph.

Fig.~\ref{fig:expansion} shows the expansion graph for making \texttt{ahu\_1} violate the constraint $\varphi$ in Fig.~\ref{fig:ahu_schema}. We write $\operatorname{VIO}(x,\varphi)$ for the  task of making node $x$ violate constraint  $\varphi$. One way to accomplish this is to remove the \texttt{hasPoint} edge. Alternatively, we can make \texttt{RAT\_Sensor} fail the  \texttt{NamedPoint} requirement, denoted  VIO(\texttt{RAT\_Sensor},\texttt{NamedPoint}), which is rewritten as VIO(\texttt{RAT\_Sensor},
        {$\exists^{\geq 1}$}\,
        \texttt{description}.
        ($\operatorname{len}\geq 3$)). This second
alternative can in turn be achieved by removing the description edge or making the ``\texttt{RAT\_01}'' string violate $\operatorname{len}\geq 3$, which is done by replacing ``RAT\_01'' with a string shorter than three characters, such as ``RA''. Details of how to construct the expansion graph are in \cite{lin2025systematic}.

The traversal of the expansion graph from the root node to a leaf node terminates in a
\textbf{normalized violation expression}: the edit operation that injects the requested violation. In this example, the three leaves yield expressions that delete the \texttt{hasPoint} edge, delete the \texttt{description} edge, or replace \texttt{"RAT\_01"} with a string such as \texttt{"RA"}. Choosing any of the leaf nodes will achieve the top level task of making \texttt{ahu\_1} violate the constraint $\varphi$. Reversing the selected operation provides the corresponding ground-truth repair.

Conceptually, we define the sampling universe $\mathcal{U}$ as the set of distinct normalized violation expressions represented by the leaves of the expansion graph.
The coverage result of \cite{lin2025systematic} establishes that $\mathcal{U}$ contains every normalized violation expression admitted by the formalism.
Given a budget $K$, our objective is to select a conflict-free subset $S\subseteq\mathcal{U}$ with $|S|=K$.
A single expression may trigger multiple validation results.
Two expressions conflict when their operations impose mutually incompatible violations on the same node.
For example, a node $x$ cannot simultaneously be perturbed to violate both $x \leq 1$ and $x > 0$.
We therefore admit only subsets whose expressions can be applied together.

The expansion graph in Fig.~\ref{fig:expansion} is small enough to show in full, but expansion graphs generally have many branches since they enumerate all violation strategies using the instance graph and schema.
Our implementation therefore does not materialize the full expansion graph to enumerate $\mathcal{U}$ before sampling.
Instead, it traverses the expansion graph incrementally, branches on the fly, and emits an expression upon reaching a leaf.
In addition, different traversal paths may yield the same normalized expression, so the sampler deduplicates and processes this stream as leaves are discovered.

\subsection{Controlling Predicate Distribution}\label{sec:pred_dist}

We treat the incrementally emitted expressions as a data stream and maintain a deduplicated reservoir of $M$ candidate expressions, from which test cases are composed.
The reservoir determines the available candidates; the final predicate distribution also depends on which candidates can be composed together.

We influence candidate selection through a positive weight $\omega_p$ assigned to each predicate $p$. We say an expression targets a predicate when that predicate is part of the edit to induce a violation. In Fig.~\ref{fig:expansion}, removing the \texttt{hasPoint} edge targets \texttt{hasPoint}, whereas removing the \texttt{description} or shortening its value targets \texttt{description}. Although all three expressions make \texttt{ahu\_1} violate the same top-level constraint, they do so through different predicates. An expression can also target multiple predicates. The weight $w(v)$ of an expression combines the weights of the predicates it targets, the geometric mean when there are several. Increasing $\omega_p$ therefore makes expressions targeting $p$ more likely to be retained in the reservoir. For the suites in Sec.~\ref{sec:hetero}, we assign larger weights to predicates in the targeted class. For router training, we estimate predicate frequencies and use inverse-frequency weights to encourage broad coverage of targeted predicates in the reservoir. 

The stream of normalized violation expressions has unknown length and is produced incrementally. We thus use the weighted reservoir sampling method of Efraimidis and Spirakis~\cite{efraimidis2006weighted}, which converts each expression's weight into a sort key. When an expression $v$ is emitted, it receives the key

\begin{equation}
    \operatorname{key}(v)
    =
    \frac{-\ln h(v)}{w(v)},
    \qquad
    h(v)=\operatorname{hash}(v)\in(0,1].
    \label{eq:reservoir-key}
\end{equation}

The reservoir retains the $M$ distinct expressions with the smallest keys. The hash assigns each expression a stable pseudorandom value, so repeated occurrences receive the same key and are counted only once. Dividing by $w(v)$ makes a larger-weight expression more likely to have a small key and thus to remain in the reservoir.

At any point, the reservoir represents a weighted sample of the
expressions discovered so far. Thus, case construction can begin
without waiting for the complete expansion graph to be traversed.
If traversal completes, the reservoir is sampled from the full
universe $\mathcal{U}$; otherwise, it represents the explored
portion of that universe.

\subsection{Composing Test Cases}

The final step composes a test case from the size-$M$ reservoir, whose candidates are selected using the predicate weights $\omega_p$.
Given a budget $K$, we select up to $K$ expressions, admitting an expression only if its edits do not conflict with those already selected. Specifically, we scan the reservoir and select an expression only if its associated violation requirements are compatible with those of the expressions already selected. We check these requirements pairwise, treating requirements on the same node as conflicting when they demand incompatible violations. Selection stops when $K$ expressions have been chosen or the reservoir is exhausted. 

We then apply the selected edits to the original graph to obtain the test case, and record their reversal as the ground-truth repair.
Separating sampling from composition also lets us draw many test cases from a single traversal of the expansion graph.

\section{Learning to Route}

\label{sec:router}

The variation in researcher utility across violations and deployments motivates learning when to enable access.
We use the test-case generator of Sec.~\ref{sec:multi_VE} to construct training cases with known reference repairs, and learn a routing policy from the observed repair outcomes.

The router operates within the iterative repair loop (Fig.~\ref{fig:repair_loop}). Immediately after the clustering algorithm groups a set of violations, the router determines the prompt configuration for the repair agent. It makes a binary choice: provide the agent with the researcher-enabled prompt, or the researcher-disabled prompt.

The router takes as input a multi-hot vector of predicate names reported for the cluster’s violations (\texttt{sh:resultPath} in SHACL); for nested constraints, we do not traverse to collect predicates from referenced shapes. The predicate grouping labels used to organize the motivating study are not included in this feature vector.

We train two XGBoost models~\cite{chen2016xgboost}: one predicts the $F_1$ score of the cluster's repair with the researcher enabled, and the other with it disabled. This two-model design is known as a T-learner in the causal inference literature~\cite{kunzel2019metalearners}; here, we use it as a predictive routing rule. At test time, the router enables the researcher when the predicted ON score exceeds the predicted OFF score. We choose XGBoost because it (a) supports the sparse, binary features and (b) enables SHAP-based interpretation of the predictions (Sec.~\ref{sec:shap}).

We train a separate router for each dataset, sampling training test cases across the full predicate set using the inverse-frequency weighting from Sec.~\ref{sec:pred_dist}. We then run each test case twice: once with the researcher always enabled, and once with it always disabled. For each cluster encountered in a trajectory, we extract its feature vector and use the recorded injected perturbations to identify its reference repair edits and compute its training $F_1$. Additional edits that do not match the reference repair count as false positives. The enabled trajectories supply training examples for the ON model, and the disabled trajectories supply examples for the OFF model.

Although ON and OFF are run on the same test case, their repair trajectories can produce different clusters. Each model learns from the clusters encountered under its corresponding policy. We evaluate the resulting router using final test-case repair quality and token consumption (Sec.~\ref{sec:exp}).



\section{Evaluation of Routing Policies}

\label{sec:exp}

We compare the T-learner router (\textbf{T-learner}) against four baselines, holding the repair loop and graph tools fixed.

\begin{itemize}[leftmargin=*]

\item {OFF} and {ON:} The baselines established in Sec.~\ref{sec:hetero}, representing always disabled and enabled retrieval, respectively. OFF serves as the reference point for repair accuracy and cost.


\item{Random:} Enables the researcher randomly with a probability set to the T-learner's provision rate for the same dataset and predicate mixture (Table~\ref{tab:main}). This gives six enabling probabilities, one for each evaluation suite. The baseline compares learned selection with random access at a matched enabling probability. Because the policies can produce different repair trajectories and cluster counts, they need not enable the same number of clusters or consume the same number of tokens.

\item {LLM Router:} A zero-shot alternative to learning the routing rule. We prompt an LLM with the same features given to T-learner, the predicates involved in the cluster, and ask whether providing the researcher will improve the repair.

\end{itemize}

\begin{figure*}
    \centering
    \includegraphics[width=\linewidth]{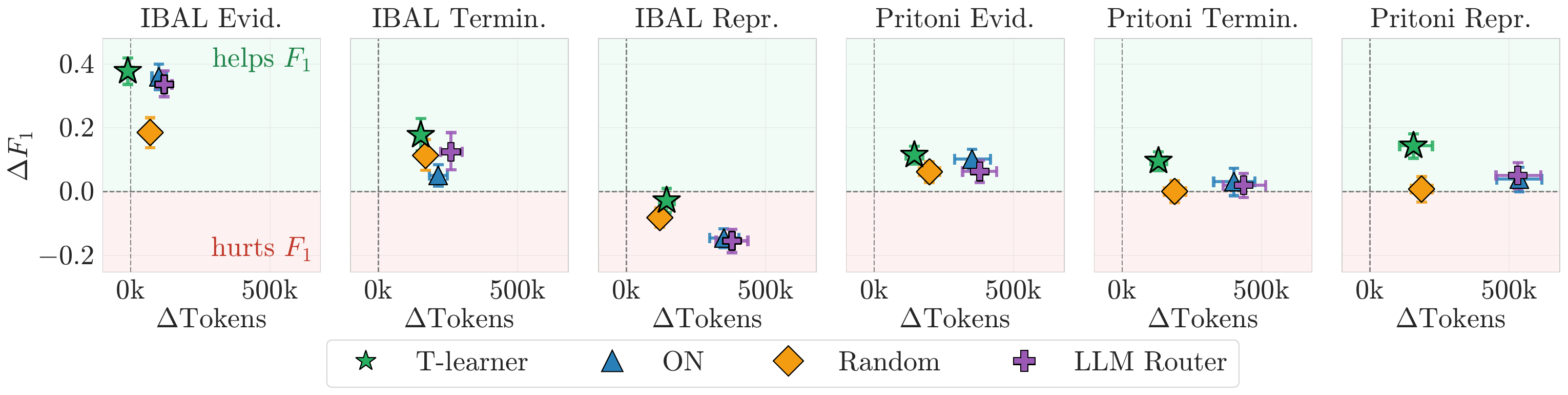}
    \caption{\normalfont\textbf{\textbf{Repair quality–token cost trade-off across datasets and predicate mixtures.}}~Each point shows the change in \(F_1\) relative to OFF (\(\Delta F_1 = F_{1,\mathrm{policy}}-F_{1,\mathrm{OFF}}\)) versus the additional token cost. The OFF \(F_1\) scores and token costs, as reported in Fig.~\ref{fig:hetero}, are $0.488$, $0.618$, $0.639$, $356\text{k}$, $245\text{k}$, and $296\text{k}$ for IBAL evidential, terminological, and representational mixtures, respectively, and $0.663$, $0.627$, $0.595$, $355\text{k}$, $379\text{k}$, and $426\text{k}$ for the corresponding Pritoni mixtures.}
    \label{fig:tradeoff}
\end{figure*}

We evaluate all five policies on the six evaluation suites detailed in Sec.~\ref{sec:hetero}. Their predicate mixtures define the evaluation workloads; the T-learner's reported provision rates are measured from its decisions on these workloads. The T-learner is trained on a disjoint set of generated test cases sampled across the full predicate set under inverse-frequency weighting, keeping evaluation cases out of router training. We use the evaluation protocol of Sec.~\ref{sec:hetero}, including its significance-testing procedure, and report the median paired difference (with a 95\% bootstrap confidence interval) in repair quality ($\Delta F_1$) and repair-time token consumption ($\Delta\text{tokens}$) against the OFF baseline for each test case (Fig.~\ref{fig:tradeoff}). The only modification is that the token counts of LLM Router include its own inference cost.

Across all 2{,}400 runs (80 test cases $\times$ 3 mixtures $\times$ 2 datasets $\times$ 5 policies), only 17 fail to produce a schema-valid graph, a failure rate below 1\%; 6 of these are the ON/OFF failures reported in Sec.~\ref{sec:hetero}. All five policies produce repairs at a high rate of schema validity, so the comparisons below concern which edits were made.

\textbf{T-learner improves $F_1$ over OFF while using fewer tokens than ON.}
In all six evaluation suites, the T-learner uses fewer tokens than ON, as shown in Fig.~\ref{fig:tradeoff}, where the T-learner lies consistently to the left of ON.
In five of them, the T-learner also improves $F_1$ relative to OFF.
The comparisons reported in Table~\ref{tab:main} indicate higher $F_1$ than ON in four of the six suites, with no statistically significant difference detected in the two evidential suites (IBAL 42/23/15, Pritoni 39/27/14). 

\textbf{Selective access on the evidential suites.}
On the two evidential suites, the T-learner enables the researcher for only 53\% of IBAL clusters and 62\% of Pritoni clusters. The comparisons detect no statistically significant $F_1$ difference from ON.

The same router is used across all evaluation suites for a dataset. Its provision rate varies with the clusters encountered in each workload; on IBAL, the rates range from 37\% to 74\% (Table~\ref{tab:main}). Sec.~\ref{sec:shap} examines how individual predicates contribute to its predictions.

\textbf{The LLM Router almost always enables researcher access.}
With the evaluated prompt and model, the LLM router enables the researcher on 99--100\% of clusters across every suite in Table~\ref{tab:main}.
It therefore provides little selectivity while adding an LLM call per cluster.
In Fig.~\ref{fig:tradeoff}, its token consumption is at or slightly above ON in almost all panels.

\textbf{The T-learner outperforms Random.}
Random enables researcher access with the same probability as the T-learner for each dataset and predicate mixture. The T-learner achieves higher repair $F_1$ in all six evaluation suites (Table~\ref{tab:main}). These results support using cluster predicate features to select researcher access.

\textbf{Selective routing mitigates harmful retrieval.}
The router improves repair $F_1$ over OFF in five of six suites.
In the remaining IBAL representational suite, where ON degrades repair accuracy, the router lowers its provision rate to 37\% and achieves higher $F_1$ with fewer tokens than ON (Fig.~\ref{fig:tradeoff}).
Selective access therefore limits the downside of harmful retrieval, although the policy remains less accurate and more expensive than OFF.

The current routing rule enables the researcher whenever
$\hat{F}_{1,\text{ON}} > \hat{F}_{1,\text{OFF}}$,
regardless of the magnitude of the predicted gain; for example, a gain of $10^{-3}$ would yield the same access decision as a gain of $0.4$.
Furthermore, the rule considers only repair accuracy, ignoring the additional token cost.
Future work will consider extensions that require a minimum predicted gain to justify researcher access or incorporate token cost and prediction uncertainty into the decision.

Clustering also determines the granularity of researcher access. Larger clusters may combine violations with different evidence needs under a single access decision. Future work will examine how clustering and routing interact, including whether adapting cluster boundaries to predicted retrieval benefit can improve repair accuracy and token efficiency.

\begin{table}[t]
\centering
\small
\setlength{\tabcolsep}{3pt}
\begin{tabular}{ll ccc cc}
\toprule
 & & \multicolumn{3}{c}{T-learner vs.\ (W/L/T)} & \multicolumn{2}{c}{Provision Rate (\%)} \\
\cmidrule(lr){3-5} \cmidrule(lr){6-7}
Dataset & Mixture &  ON & Random & LLM & LLM & T-learner \\
\midrule
IBAL & Evid. & 42/23/15$^{\text{ns}}$ & 66/5/9$^{\text{***}}$ & 46/15/19$^{\text{**}}$ & 99 & 53 \\
IBAL & Termin. & 57/17/6$^{\text{***}}$ & 53/21/6$^{\text{***}}$ & 48/21/11$^{\text{**}}$ & 99 & 74 \\
IBAL & Repr. & 61/12/7$^{\text{***}}$ & 44/27/9$^{\text{*}}$ & 59/15/6$^{\text{***}}$ & 99 & 37 \\
\midrule
Pritoni & Evid. & 39/27/14$^{\text{ns}}$ & 46/25/9$^{\text{**}}$ & 43/20/17$^{\text{**}}$ & 99 & 62 \\
Pritoni & Termin. & 51/21/8$^{\text{***}}$ & 56/16/8$^{\text{***}}$ & 56/15/9$^{\text{***}}$ & 100 & 57 \\
Pritoni & Repr. & 61/17/2$^{\text{***}}$ & 68/9/3$^{\text{***}}$ & 62/14/4$^{\text{***}}$ & 99 & 52 \\
\bottomrule
\end{tabular}
\caption{\normalfont\textbf{Repair quality and provision rate across datasets and predicate mixtures.}~W/L/T counts test cases in which the T-learner attains higher / lower / equal $F_1$ than the competitor; reported significance markers are $^{***}p<.001$, $^{**}p<.01$, $^{*}p<.05$, and ns $=$ not significant. 
}\label{tab:main}
\vspace{-2em}
\end{table}

\section{Interpreting the T-learner} \label{sec:shap}



\begin{figure}
    \centering
    \includegraphics[width=\linewidth]{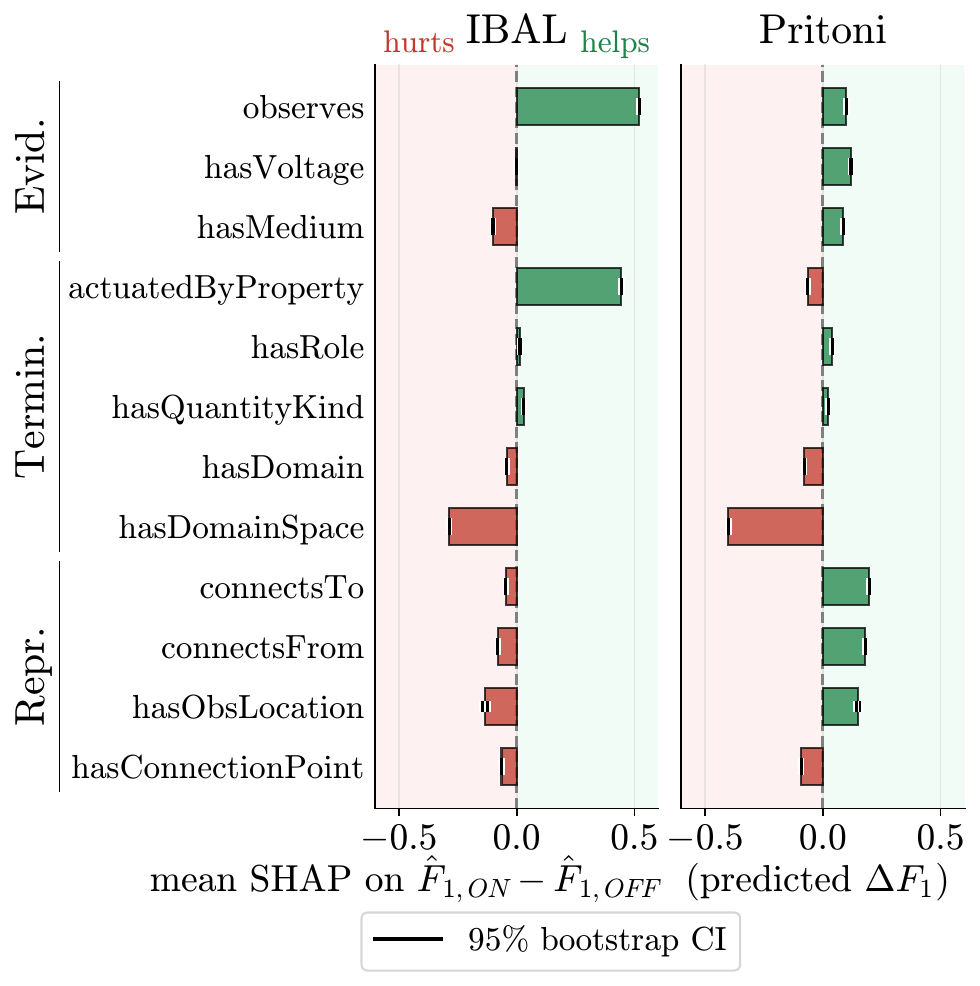}
    \caption{\textbf{Predicate-level SHAP values for the T-learner's predicted retrieval gain.}~The plot's ``helps'' and ``hurts'' labels refer to contributions that raise or lower predicted retrieval gain relative to the base value. The reported base values are $+0.1602$ for IBAL and $+0.0527$ for Pritoni.}\label{fig:shap}
    \vspace{-3em}
\end{figure}

Sec.~\ref{sec:exp} shows that the T-learner selectively enables researcher access across the evaluation suites. 
Fig.~\ref{fig:shap} examines how individual predicates contribute to the router's predicted gain, $\hat{F}_{1,\mathrm{ON}}-\hat{F}_{1,\mathrm{OFF}}$, using SHAP~\cite{lundberg2017unified}. We compute contributions separately for the ON and OFF models and subtract them; positive contributions raise the predicted gain relative to its base value, while negative contributions lower it. Each bar shows the mean contribution across clusters from that deployment's training mixtures in which the predicate is present. Whiskers show 95\% percentile bootstrap intervals for that mean, based on 5,000 resamples. The router enables researcher access when the total predicted gain is positive.


The learned contributions differ among predicates, including those within the same class. For example, on IBAL, \texttt{observes} has a positive mean contribution to predicted gain, whereas \texttt{hasMedium} has a negative one, although both belong to the evidential class. The same predicate can also contribute differently across deployments: the mean contribution of \texttt{hasMedium} is positive on Pritoni. These patterns illustrate the finer distinctions learned from repair outcomes within each deployment.


SHAP decomposes each cluster's predicted gain into a base value plus the contributions of its predicate features. On IBAL, this base value is $+0.1602$, roughly three times that on Pritoni ($+0.0527$), indicating a higher baseline predicted benefit from researcher access. The predicate contributions must therefore offset a larger positive baseline before the IBAL router disables access. This helps interpret how access can remain enabled even when individual predicates, such as \texttt{hasConnectionPoint}, contribute negatively to predicted retrieval gain. Each access decision depends on the combined contributions for an individual cluster, while Fig.~\ref{fig:shap} summarizes mean contributions across clusters.


Overall, these patterns help interpret how the router selects researcher access using individual predicate features. The evaluation suites measure the repair quality and token consumption of applying the same learned policy across different predicate mixtures within each deployment.

\section{Limitations \& Discussion}\label{sec:limit}

Our evaluation is limited to two building-automation deployments using the 223P schema. This scope reflects a benchmark constraint: as discussed in Sec.~\ref{sec:datasets}, few datasets provide a schema, a valid instance graph, and an external corpus simultaneously. Widely used knowledge bases also risk confounding retrieval evaluation through contamination of the LLM's parametric memory. The contrast between Pritoni and IBAL demonstrates that retrieval utility varies even within one domain and schema, but it does not establish generalization to other domains. Evaluating such settings will require additional benchmarks with the same three aligned artifacts.

Training the router requires a trusted reference graph and an aligned document corpus for the deployment. Its applicability therefore depends on the availability and coverage of that reference. An existing graph may contain a trusted, schema-valid subgraph from which our multi-violation generation framework (Sec.~\ref{sec:multi_VE}) can construct training cases with known reference repairs. Running these cases with researcher access enabled and disabled supplies outcomes for training the router. Repeating this procedure could support periodic retraining as the graph, corpus, or retrieval pipeline changes. Preparing the corpus index, generating and labeling cases, and training the router also incur costs beyond repair-time token consumption.

Furthermore, our analysis is also limited in its ability to attribute retrieval failures. We treat the external corpus and the researcher as a unified retrieval pipeline because the repair agent observes only the researcher's final report. When this pipeline fails to supply a required fact, we cannot distinguish among the following scenarios:

\begin{enumerate}[leftmargin=*]

\item The fact is explicitly present in the external corpus, but the researcher fails to retrieve or extract it.

\item The fact is not explicitly stated but can be inferred by combining evidence across documents, and the researcher fails to make this inference.

\item Neither the fact nor sufficient evidence from which to infer it is present in the external corpus.

\end{enumerate}

These scenarios are observationally equivalent to the repair agent because each results in a report that does not provide the required information. Disentangling corpus coverage from researcher efficacy would require ground-truth annotations. Constructing this ground truth is beyond the scope of this paper.

Finally, the T-learner router is trained offline and held fixed so it does not adapt to deployment drift such as new document additions to the corpus or changes to the retrieval pipeline. Online adaptation with feedback using methods such as contextual bandits~\cite{li2010contextual} remains future work.

\section{Conclusion}

This paper studies when an LLM-based graph repair agent should be given access to an external retrieval system.
Our experiments show that the effect of researcher access differs across repairs and corpora: it can improve repair accuracy, increase token consumption without improving accuracy, or even lead to less accurate repairs.
This variation motivates learning a deployment-specific access policy.

We therefore learn a deployment-specific router from automatically generated repair cases. Our test-generation framework composes multiple compatible constraint violations while retaining programmatic ground-truth repairs and controlling predicate distribution. Using outcomes collected with the researcher enabled and disabled, the router learns predicate-level differences in retrieval utility and selectively provisions researcher access for each violation cluster. Across six evaluation suites on two building-automation deployments, the router uses fewer repair-time tokens than always enabling access in all six suites. It improves repair $F_1$ over always enabling access in four suites and over always disabling access in five. These results demonstrate that learned, selective access to external retrieval can improve repair accuracy while reducing token consumption.

\bibliographystyle{ACM-Reference-Format}
\bibliography{sample}

\end{document}